\documentclass[preprint,12pt]{elsarticle}

\usepackage{graphicx}
\usepackage{subfigure}
\usepackage{amsmath}
\usepackage{amsfonts}
\usepackage{amssymb}
\usepackage{endfloat}

\journal{Physica A}

\begin{document}

\begin{frontmatter}

\title{Critical and tricritical behavior of a selectively diluted triangular Ising antiferromagnet in a field}
\author{M. Borovsk\'{y}}
%\ead{borovsky.michal@gmail.com}
\author{M. \v{Z}ukovi\v{c}\corref{cor1}}
\ead{milan.zukovic@upjs.sk}
\author{A. Bob\'ak}
%\ead{andrej.bobak@upjs.sk}
\address{Department of Theoretical Physics and Astrophysics, Faculty of Science,\\ 
P.J. \v{S}af\'arik University, Park Angelinum 9, 041 54 Ko\v{s}ice, Slovak Republic}
\cortext[cor1]{Corresponding author}

\begin{abstract}
We study a geometrically frustrated triangular Ising antiferromagnet in an external magnetic field which is selectively diluted with nonmagnetic impurities employing an effective-field theory with correlations and Monte Carlo simulations. We focus on the frustration-relieving effects of such a selective dilution on the phase diagram and find that it can lead to rather intricate phase diagrams in the dilution-field parameters space. In particular, in a highly (weakly) diluted system the frustration is greatly (little) relieved and such a system is found to display only the second(first)-order phase transitions at any field. On the other hand, for a wide interval of intermediate dilution values the transition remains second order at low fields but it changes to first order at higher fields and the system displays a tricritical behavior. The existence of the first-order transition in the region of intermediate dilution and high fields is verified by Monte Carlo simulations.
\end{abstract}


\end{frontmatter}

\section{Introduction}
A triangular Ising antiferromagnet is a typical geometrically frustrated spin model which due to a high degree of frustration shows no long-range ordering~\cite{wannier,hout}. Nevertheless, a small perturbation, for example in the form of an applied external magnetic field or the injection of quenched non-magnetic impurities, can lift the high degeneracy of the ground state and thus initiate long-range ordering. In particular, it has been shown that for a certain range of the field values the systems displays a transition between the ferrimagnetic phase with two sublattices aligned parallel and one antiparallel to the field at lower temperatures and the paramagnetic phase in which all spins are aligned parallel to the field at higher temperatures~\cite{metcalf,schick,netz}. On the other hand, the injection of quenched magnetic vacancies in zero field can relieve frustration but only locally, which supposedly leads to a spin-glass order~\cite{grest,ander,blac}. Combining both the presence of the external field and the magnetic impurities can lead to rather interesting phenomena, such as step-wise field-dependence of the magnetization or possibility of multiple reentrants in the phase diagram, as suggested by recent Monte Carlo~\cite{yao} and effective-field theory~\cite{zuko} studies. Another way of relieving the geometrical frustration is a selective dilution, in which different (in our case three) sublattices are diluted by non-magnetic impurities with different probabilities. Then the geometrical frustration can also be relieved globally, giving rise to long-range magnetic ordering phenomena even in the absence of the field. Kaya and Berker~\cite{kaya} have shown that if only one out of three sublattices of the triangular lattice is randomly diluted then a long-range order can develop in the remaining two sublattices already at relatively low concentrations of the vacancies. Various studies have shown that the presence of either the external magnetic field or the magnetic impurities has generally different effects on the critical behavior of non-frustrated and frustrated spin systems. For example, in the non-frustrated spin systems the critical temperature typically decreases with the field (e.g.,~\cite{kinc,wu,wang}) and in some peculiar systems, such as metamagnets~\cite{kinc}, the transition in a high-field and low-temperature region can change to a first-order one. On the other hand, in the frustrated systems the magnetic field can either enhance or suppress the transition temperature, depending on the field range~\cite{metcalf,schick,netz,zuko}. Similarly, the introduction of the quenched dilution in the non-frustrated systems typically decreases the transition temperature down to zero at the percolation threshold~\cite{stin}, while the opposite effect also known as ``order by disorder'' can be achieved by a selective quenched dilution in the frustrated systems~\cite{kaya}. Therefore, by applying the external field to the selectively diluted system we can anticipate different regimes of its effects on the critical properties of the system and thus an interesting phase diagram.\\
\hspace*{5mm} The goal of the present study is to investigate changes in the critical behavior of the system, the frustration of which is systematically controlled by the degree of the selective dilution between the fully frustrated (pure system on triangular lattice) and non-frustrated (pure system on honeycomb lattice) limits.

\section{The model and method}
\begin{figure}[t!]
\centering
    \includegraphics[scale=.5]{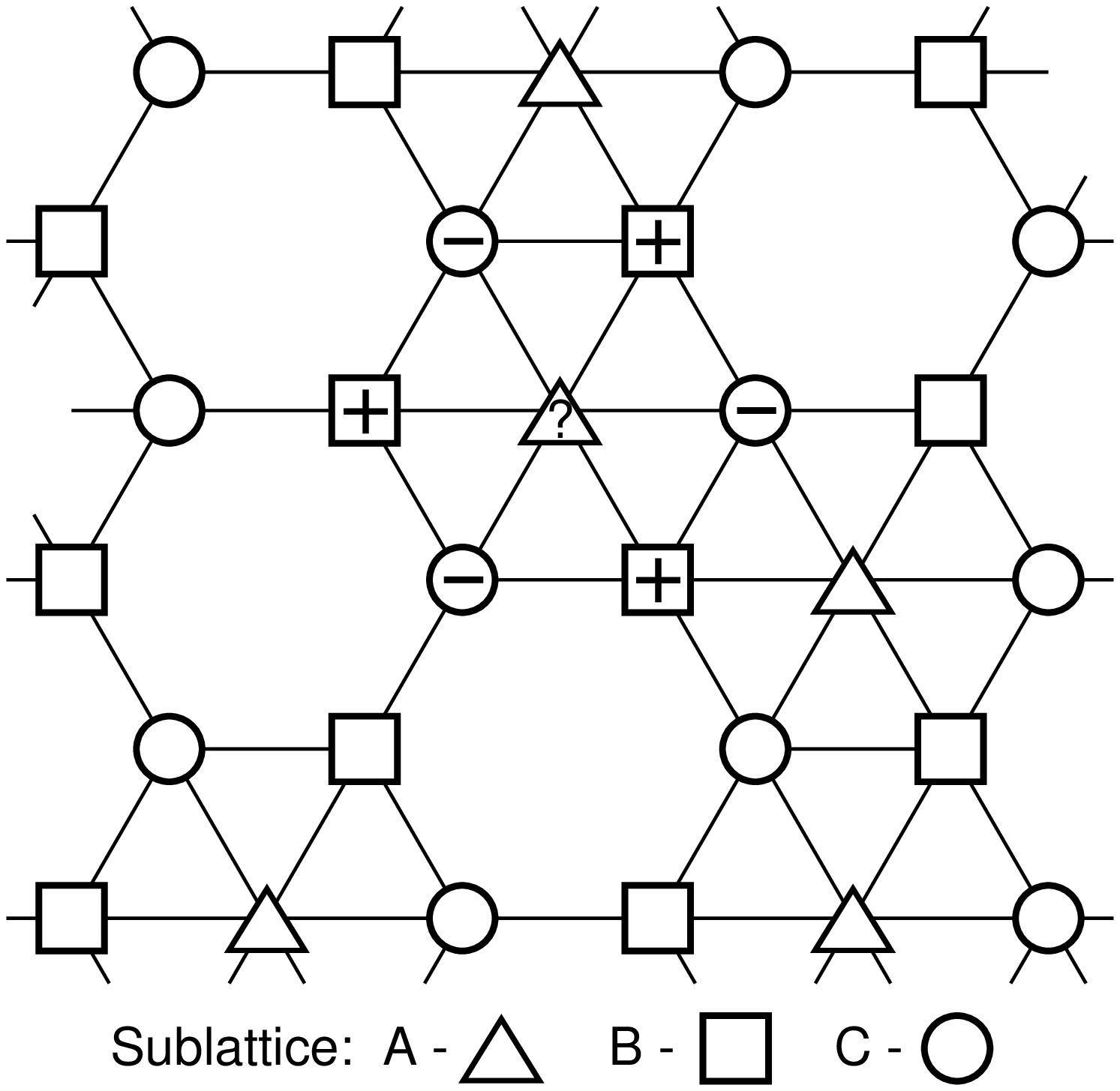}
\caption{Triangular lattice partition into three sublattices A,B and C, where the sublattice A is randomly diluted. The signs $+$ and $-$ represent the spin values of $+1$ and $-1$, respectively, and the question mark signifies the frustration arising from placing a magnetic A-ion in the center of the antiferromagnetically ordered hexagon of B- and C-ions.}
\label{fig:lattice}
\end{figure}

The selectively diluted Ising antiferromagnet in a field can be described by the Hamiltonian
\begin{equation}
\label{Hamiltonian}
H=-J\sum_{\langle i,j \rangle}\xi_{i}\xi_{j}S_{i}S_{j}-h\sum_{i}\xi_{i}S_{i},
\end{equation}
where $S_{i}=\pm1$, are the Ising spins, $h$ is the external magnetic field, $J<0$ is the exchange interaction parameter, and $\langle i,j \rangle$ denotes the summation over all nearest neighbor (NN) pairs. $\xi_{i}$ are quenched, uncorrelated random variables which are equal to $1$ with probability $p$ when the site $i$ is occupied by a magnetic atom and $0$ with probability $1-p$ otherwise. Then $p$ represents the mean concentration of magnetic atoms. For the current selectively diluted case with only one (let us say A) sublattice diluted we will consider the sublattice-dependent concentrations $0 \leq p_{\mathrm{A}} \leq 1$, $p_{\mathrm{B}} \equiv p_{\mathrm{C}}=1$. \\  
\hspace*{5mm} In order to study the critical behavior of the system, we apply an effective field theory with correlations (EFT) (for review see, e.g.,~\cite{kane}). In contrast to the usual mean-field approach, EFT correctly accounts for the single-site kinematic relations through the Van der Waerden identity. As a result, EFT yields, for example, a non-zero critical concentration for quenched diluted systems, the lack of order in the one-dimensional Ising ferromagnet and the occurrence of order in the two-dimensional case with the critical temperature improved over the usual mean-field theory~\cite{kane}. However, an extra care must be taken when dealing with a frustrated system since a straightforward application of EFT may lead a complete loss of frustration and consequently meaningless results. Recently, we have applied EFT to an uniformly diluted triangular Ising antiferromagnet~\cite{zuko} and in the present study we will follow the same procedure in order to include the geometrical frustration effects within EFT. Namely, we decompose the triangular lattice into three interpenetrating sublattices A, B and C (see Fig.~\ref{fig:lattice}), such a way that spins on one sublattice can only interact with their NNs on the other two sublattices. Thus all the NN interactions are accounted for and the frustration results from the effort to simultaneously satisfy all the mutual antiferromagnetic intersublattice couplings.  
Such an approach has correctly reproduced no long-range order behavior down to zero temperature for the pure system in zero field~\cite{zuko}.

\subsection{Sublattice magnetizations}
The approach addopted in Ref.~\cite{zuko} can be generalized by considering the concentration $p$ sublattice-dependent and hence characterized by a three-component vector ${\bf p}=(p_{\mathrm{A}},p_{\mathrm{B}},p_{\mathrm{C}})$. Then, the A-sublattice magnetization per site can be calculated by taking a configurational average of the expression
\begin{equation}
\label{conf_sub}
\begin{array}{l}
\xi_{i}^{\mathrm{A}}\langle S_{i}^{\mathrm{A}}\rangle=\xi_{i}^{\mathrm{A}}\Big\langle
\prod_{j=1}^{z_{\mathrm{AB}}}[\xi_{j}^{\mathrm{B}}\cosh(\beta JD)+\xi_{j}^{\mathrm{B}}S_{j}^{\mathrm{B}}\sinh(\beta JD)+1-\xi_{j}^{\mathrm{B}}]\\ 
\times\prod_{k=1}^{z_{\mathrm{AC}}}[\xi_{k}^{\mathrm{C}}\cosh(\beta JD)+\xi_{k}^{\mathrm{C}}S_{k}^{\mathrm{C}}\sinh(\beta JD)+1-\xi_{k}^{\mathrm{C}}]\Big\rangle\tanh(x+\beta h)|_{x=0},
\end{array}
\end{equation}
where $z_{\mathrm{AB}},z_{\mathrm{AC}}$ are the numbers of NNs of the spin $S_{i}^{\mathrm{A}}$ from the sublattice A that belong to the sublattices B and C, respectively, $\beta=1/k_{B}T$ and $D=\partial/\partial x$ is the differential operator. We can obtain similar expressions for the sublattices B and C from Eq.~\eqref{conf_sub} by cyclic permutation of the indices A, B and C and for each sublattice taking an appropriate number of NNs on the remaining sublattices (in the present case $z_{\mathrm{XY}}=3$, for all pairs X,Y $\in {\mathrm{\{A,B,C}\}}$). Then, we apply the decoupling approximation for the thermal averaging~\cite{zuko} and perform the configurational averaging of these equations for the respective sublattices. For the selectively diluted case considered in the present paper, i.e., $\left( p_{\mathrm{A}}, p_{\mathrm{B}}, p_{\mathrm{C}} \right) = \left( p_{\mathrm{A}}, 1, 1 \right)$, this leads to a system of coupled equations for the averaged sublattice magnetizations per site in the form

\begin{equation}
\begin{array}{l}
		\label{sub_mag_sel}
				m_{\mathrm{A}} = p_{\mathrm{A}} \left( a + b m_{\mathrm{B}} \right)^3 \left( a + b m_{\mathrm{C}} \right)^3 \tanh \left( x + \beta h \right) |_{x=0}, \\
				m_{\mathrm{B}} = \left( a_{\mathrm{A}} + b m_{\mathrm{A}} \right)^3 \left( a + b m_{\mathrm{C}} \right)^3 \tanh \left( x + \beta h \right) |_{x=0}, \\
				m_{\mathrm{C}} = \left( a_{\mathrm{A}} + b m_{\mathrm{A}} \right)^3 \left( a + b m_{\mathrm{B}} \right)^3 \tanh \left( x + \beta h \right) |_{x=0},
\end{array}
\end{equation}
where $a = \cosh \left( \beta J D \right)$, $b = \sinh \left( \beta J D \right)$ and $a_{\mathrm{A}} = 1 - p_{\mathrm{A}} + p_{\mathrm{A}} \cosh \left( \beta J D \right)$. By expanding the right-hand sides of Eqs.~\eqref{sub_mag_sel} we obtain

\begin{equation}
\begin{array}{r l}
		\label{sub_mag_honey2}
				m_{\mathrm{A}} = &K_0 + K_1 \left( m_{\mathrm{B}} + m_{\mathrm{C}} \right) + 
				K_2 \left( m_{\mathrm{B}}^2 + 3 m_{\mathrm{B}} m_{\mathrm{C}} + m_{\mathrm{C}}^2 \right) + \\
				& K_3 \left( m_{\mathrm{B}}^3 + 9 m_{\mathrm{B}}^2 m_{\mathrm{C}} 
				+ 9 m_{\mathrm{B}} m_{\mathrm{C}}^2 + m_{\mathrm{C}}^3 \right) + 	\\			 
				& K_4 \left( m_{\mathrm{B}}^3 m_{\mathrm{C}} + 3 m_{\mathrm{B}}^2 m_{\mathrm{C}}^2  
				+ m_{\mathrm{B}} m_{\mathrm{C}}^3 \right) + \\
				& K_5 \left( m_{\mathrm{B}}^3 m_{\mathrm{C}}^2  + m_{\mathrm{B}}^2 m_{\mathrm{C}}^3 \right) 
				+ K_6 m_{\mathrm{B}}^3 m_{\mathrm{C}}^3 , \\
				
				m_{\mathrm{B}} = & L_0 + L_1 m_{\mathrm{A}} + L_2 m_{\mathrm{C}} 
				+ L_3 m_{\mathrm{A}}^2 + L_4 m_{\mathrm{C}}^2 + L_5 m_{\mathrm{A}} m_{\mathrm{C}} + \\
				& L_6 m_{\mathrm{A}}^3  + L_7 m_{\mathrm{C}}^3 
				+ L_8 m_{\mathrm{A}}^2 m_{\mathrm{C}} + L_9 m_{\mathrm{A}} m_{\mathrm{C}}^2 + \\
				& L_{10} m_{\mathrm{A}}^2 m_{\mathrm{C}}^2 + L_{11} m_{\mathrm{A}}^3 m_{\mathrm{C}} +
				L_{12} m_{\mathrm{A}} m_{\mathrm{C}}^3 \\
				& L_{13} m_{\mathrm{A}}^3 m_{\mathrm{C}}^2 + L_{14} m_{\mathrm{A}}^2 m_{\mathrm{C}}^3 +				
				L_{15} m_{\mathrm{A}}^3 m_{\mathrm{C}}^3, \\
				
				m_{\mathrm{C}} = & L_0 + L_1 m_{\mathrm{A}} + L_2 m_{\mathrm{B}} 
				+ L_3 m_{\mathrm{A}}^2 + L_4 m_{\mathrm{B}}^2 + L_5 m_{\mathrm{A}} m_{\mathrm{B}} + \\
				& L_6 m_{\mathrm{A}}^3  + L_7 m_{\mathrm{B}}^3 
				+ L_8 m_{\mathrm{A}}^2 m_{\mathrm{B}} + L_9 m_{\mathrm{A}} m_{\mathrm{B}}^2 + \\
				& L_{10} m_{\mathrm{A}}^2 m_{\mathrm{B}}^2 + L_{11} m_{\mathrm{A}}^3 m_{\mathrm{B}} +
				L_{12} m_{\mathrm{A}} m_{\mathrm{B}}^3 \\
				& L_{13} m_{\mathrm{A}}^3 m_{\mathrm{B}}^2 + L_{14} m_{\mathrm{A}}^2 m_{\mathrm{B}}^3 +				
				L_{15} m_{\mathrm{A}}^3 m_{\mathrm{B}}^3,
				
\end{array}
\end{equation}
where the coefficients $K_i$ ($i = 0-6$) and $L_j$ ($j = 0-15$) are given in the appendix. 

\subsection{Critical and tricritical points}
In order to determine the critical boundaries, let us introduce new variables: the mean magnetization of the non-diluted sublattices B and C, $m_{\mathrm{BC}} = \left( m_{\mathrm{B}} + m_{\mathrm{C}} \right) / 2$, and the order parameter $m_{\mathrm{O}} = \left( m_{\mathrm{B}} - m_{\mathrm{C}} \right) / 2$. The latter serves to distinguish between the ordered ferrimagnetic ($m_{\mathrm{O}}>0$) and the paramagnetic ($m_{\mathrm{O}}=0$) phases. Then, Eqs.~\eqref{sub_mag_honey2} can be rewritten in terms of $m_{\mathrm{A}}$, $m_{\mathrm{BC}}$, and $m_{\mathrm{O}}$ as follows:

\begin{equation}
\begin{array}{r l}
			m_{\mathrm{A}} = 
			& A_0\left( m_{\mathrm{BC}} \right) + A_2\left( m_{\mathrm{BC}} \right) m_{\mathrm{O}}^2 + A_4\left( m_{\mathrm{BC}} \right) m_{\mathrm{O}}^4 + A_6 m_{\mathrm{O}}^6, \\
%		\label{eq_mA}
%\end{equation}
%\begin{equation}
			m_{\mathrm{BC}} = 
			& B_0\left( m_{\mathrm{A}}, m_{\mathrm{BC}} \right) + B_2\left( m_{\mathrm{A}}, m_{\mathrm{BC}} \right) m_{\mathrm{O}}^2, \\
%		\label{eq_mBC}
%\end{equation}
%\begin{equation}
			m_{\mathrm{O}} = 
			& C_1\left( m_{\mathrm{A}}, m_{\mathrm{BC}} \right) m_{\mathrm{O}} + C_3\left( m_{\mathrm{A}}, m_{\mathrm{BC}} \right) m_{\mathrm{O}}^3,		
%		\label{eq_orderPar}
		\label{mA_mBC_o}
\end{array}
\end{equation}
where
\begin{equation}
\begin{array}{r l}
				A_0\left( m_{\mathrm{BC}} \right) = 
				& K_0 + 2 K_1 m_{\mathrm{BC}} + 5 K_2 m_{\mathrm{BC}}^2 + 20 K_3 m_{\mathrm{BC}}^3 + \\
				&  5 K_4 m_{\mathrm{BC}}^4 + 2 K_5 m_{\mathrm{BC}}^5 + K_6 m_{\mathrm{BC}}^6, \\
				A_2\left( m_{\mathrm{BC}} \right) = & - \left( K_2 + 12 K_3 m_{\mathrm{BC}} + 6 K_4 m_{\mathrm{BC}}^2 + 
				4 K_5 m_{\mathrm{BC}}^3 + 3 K_6 m_{\mathrm{BC}}^4  \right), \\
				A_4\left( m_{\mathrm{BC}} \right) = & K_4 + 2 K_5 m_{\mathrm{BC}} + 3 K_6 m_{\mathrm{BC}}^2, \\
				A_6\left( m_{\mathrm{BC}} \right) = & - K_6,				
		\label{coeffsA}
\end{array}
\end{equation}
\begin{equation}
\begin{array}{r l}
				B_0\left( m_{\mathrm{A}}, m_{\mathrm{BC}} \right) = 
				& L_0 + L_1 m_{\mathrm{A}} + L_2 m_{\mathrm{BC}} 
				+ L_3 m_{\mathrm{A}}^2 + L_4 m_{\mathrm{BC}}^2 + L_5 m_{\mathrm{A}} m_{\mathrm{BC}} + \\
				& L_6 m_{\mathrm{A}}^3  + L_7 m_{\mathrm{BC}}^3 
				+ L_8 m_{\mathrm{A}}^2 m_{\mathrm{BC}} + L_9 m_{\mathrm{A}} m_{\mathrm{BC}}^2 + \\
				& L_{10} m_{\mathrm{A}}^2 m_{\mathrm{BC}}^2 + L_{11} m_{\mathrm{A}}^3 m_{\mathrm{BC}} +
				L_{12} m_{\mathrm{A}} m_{\mathrm{BC}}^3 \\
				& L_{13} m_{\mathrm{A}}^3 m_{\mathrm{BC}}^2 + L_{14} m_{\mathrm{A}}^2 m_{\mathrm{BC}}^3 +				
				L_{15} m_{\mathrm{A}}^3 m_{\mathrm{BC}}^3, \\
				B_2\left(  m_{\mathrm{A}}, m_{\mathrm{BC}} \right) = & L_4 + 3 L_7 m_{\mathrm{BC}} + L_9 m_{\mathrm{A}} + 
				L_{10} m_{\mathrm{A}}^2 + 3 L_{12} m_{\mathrm{A}} m_{\mathrm{BC}} \\
				& L_{13} m_{\mathrm{A}}^3 + 3 L_{14} m_{\mathrm{A}}^2 m_{\mathrm{BC}} +				
				3 L_{15} m_{\mathrm{A}}^3 m_{\mathrm{BC}},			
		\label{coeffsB}
\end{array}
\end{equation}
\begin{equation}
\begin{array}{r l}
				C_1\left( m_{\mathrm{A}}, m_{\mathrm{BC}} \right) = 
				& - ( L_2 + 2 L_4 m_{\mathrm{BC}} + L_5 m_{\mathrm{A}} + 3 L_7 m_{\mathrm{BC}}^2 
				+ L_8 m_{\mathrm{A}}^2 + \\ 
				& 2 L_9 m_{\mathrm{A}} m_{\mathrm{BC}} + 2 L_{10} m_{\mathrm{A}}^2 m_{\mathrm{BC}} +
				L_{11} m_{\mathrm{A}}^3 +	3 L_{12} m_{\mathrm{A}} m_{\mathrm{BC}}^2 \\
				& 2 L_{13} m_{\mathrm{A}}^3 m_{\mathrm{BC}} + 3 L_{14} m_{\mathrm{A}}^2 m_{\mathrm{BC}}^2 +				
				3 L_{15} m_{\mathrm{A}}^3 m_{\mathrm{BC}}^2 ), \\
				C_3\left( m_{\mathrm{A}} \right) = 
				& - \left( L_7 + L_{12} m_{\mathrm{A}} + L_{14} m_{\mathrm{A}}^2 + 
				L_{15} m_{\mathrm{A}}^3 \right),
		\label{coeffsC}
\end{array}
\end{equation}

Now, by solving Eqs.~\eqref{mA_mBC_o} in the neighborhood of a second-order transition line, where $m_{\mathrm{O}}$ is small, and retaining the terms up to the third order of $m_{\mathrm{O}}$, one can obtain
\begin{equation}
			m_{\mathrm{A}} = m_{\mathrm{A}_0} + m_{\mathrm{A}_1} m_{\mathrm{O}}^2 + \hdots, 
		\label{eq_CP_mA}
\end{equation}
\begin{equation}
			m_{\mathrm{BC}} = m_{\mathrm{BC}_0} + m_{\mathrm{BC}_1} m_{\mathrm{O}}^2 + \hdots, 
		\label{eq_CP_mBC}
\end{equation}
\begin{equation}
			m_{\mathrm{O}} = \alpha m_{\mathrm{O}} + \beta m_{\mathrm{O}}^3 + \hdots,		
		\label{eq_CP_orderPar}		
\end{equation}
where $m_{\mathrm{A}_0}$ and $m_{\mathrm{BC}_0}$ are solutions of 
\begin{equation}
				m_{\mathrm{A}_0} = A_0\left( m_{\mathrm{BC}_0} \right),
		\label{mA_0}
\end{equation}
\begin{equation}
				m_{\mathrm{BC}_0} = B_0\left(  m_{\mathrm{A}_0}, m_{\mathrm{BC}_0} \right),
		\label{mBC_0}
\end{equation}
and $m_{\mathrm{A}_1}$, $m_{\mathrm{BC}_1}$, $\alpha$ and $\beta$ are given by
\begin{equation}
				m_{\mathrm{A}_1} = E_0(m_{\mathrm{BC}_0}) m_{\mathrm{BC}_1} + A_2\left( m_{\mathrm{BC}_0} \right),
		\label{mA_1}
\end{equation}				
\begin{equation}
				m_{\mathrm{BC}_1} = \frac{E_1(m_{\mathrm{A}_0},	m_{\mathrm{BC}_0}) A_2\left( m_{\mathrm{BC}_0} \right) + B_2\left(  m_{\mathrm{A}_0}, m_{\mathrm{BC}_0} \right)}{1 - E_0(m_{\mathrm{BC}_0}) E_1(m_{\mathrm{A}_0},	m_{\mathrm{BC}_0}) - E_2(m_{\mathrm{A}_0}, m_{\mathrm{BC}_0})},
		\label{mBC_1}
\end{equation}
\begin{equation}
				\alpha = C_1\left(  m_{\mathrm{A}_0}, m_{\mathrm{BC}_0} \right),
		\label{coeff_a}
\end{equation}
\begin{equation}
				\beta = C_3\left(  m_{\mathrm{A}_0} \right) + E_3(m_{\mathrm{A}_0}, m_{\mathrm{BC}_0}) m_{\mathrm{A}_1} + E_4(m_{\mathrm{A}_0}, m_{\mathrm{BC}_0}) m_{\mathrm{BC}_1},
		\label{coeff_b}
\end{equation}
where 
\begin{equation}
		E_0(m_{\mathrm{BC}_0}) = \frac{dA_0 (m_{\mathrm{BC}_0})}{dm_{\mathrm{BC}}},\ E_1(m_{\mathrm{A}_0},	m_{\mathrm{BC}_0}) = \frac{\partial B_0 (m_{\mathrm{A}_0},	m_{\mathrm{BC}_0})}{\partial m_{\mathrm{A}}}, \nonumber
		\label{coeff_E0-E1}
\end{equation}
\begin{equation}		
		E_2(m_{\mathrm{A}_0}, m_{\mathrm{BC}_0}) = \frac{\partial B_0 (m_{\mathrm{A}_0}, m_{\mathrm{BC}_0})}{\partial m_{\mathrm{BC}}},\ E_3(m_{\mathrm{A}_0}, m_{\mathrm{BC}_0}) = \frac{\partial C_1 (m_{\mathrm{A}_0}, m_{\mathrm{BC}_0})}{\partial m_{\mathrm{A}}}, \nonumber
		\label{coeff_E2-E3}
\end{equation}
\begin{equation}		
		E_4(m_{\mathrm{A}_0}, m_{\mathrm{BC}_0}) = \frac{\partial C_1 (m_{\mathrm{A}_0}, m_{\mathrm{BC}_0})}{\partial m_{\mathrm{BC}}}. \nonumber
		\label{coeff_E4}
\end{equation}

%respectively, represent the values of $m_{\mathrm{A}}$ and $m_{\mathrm{BC}}$ at the transition, and the coefficients $\alpha$ and $\beta$ generally depend on $m_{\mathrm{A}_0}$, $m_{\mathrm{BC}_0}$, $m_{\mathrm{A}_1}$, and $m_{\mathrm{BC}_1}$. 

The second-order phase transition line is then determined by $\alpha = 1$ and $\beta < 0$. In the vicinity of the second-order phase transition line, the order parameter is given by
\begin{equation}
			m_{\mathrm{O}}^2 = (1-\alpha)/\beta.	
		\label{o_sq}		
\end{equation}
The right-hand side of the Eq.~\eqref{o_sq} must be positive. If this is not the case, the transition is of the first order, and hence the point at which
\begin{equation}
\alpha = 1, \beta = 0
		\label{TCP}		
\end{equation}
is the tricritical point (see, e.g.,~\cite{zuko0} and references therein). 

\section{Results and discussion}
\subsection{Ground-state configurations}
Let us first examine the ground-state spin configurations of the selectively diluted system with $0 \le p_\mathrm{A} \le 1$ for different field values. Based on simple energy arguments it is easy to verify that, depending on the field value, there are three ground-state configurations. Their characteristics are summarized in Table \ref{tab:GS}. The cases of $p_{\mathrm{A}}=1$ and $p_{\mathrm{A}}=0$ correspond to the pure frustrated antiferromagnet on a triangular lattice and the pure non-frustrated antiferromagnet on a honeycomb lattice, respectively (see Fig.~\ref{fig:lattice}). The former three-sublattice system is known to display the ferrimagnetic phase with two sublattices aligned parallel and one antiparallel to the field within the field range $0 < h/|J| \le 6$ ~\cite{metcalf,schick,netz,zuko}, while the latter two-sublattice system shows an antiferromagnetic alignment within the field range $0 \le h/|J| \le 3$~\cite{wu,wang}. Therefore, for the field values $h/|J| \in \left(0,3\right]$, both these limiting cases display phase transitions, and we expect the phase transitions to occur also for intermediate values of $p_{\mathrm{A}}$. For $3 < h/|J| \le 6$, the system with $p_{\mathrm{A}}=0$, i.e. the antiferromagnet on a honeycomb lattice, does not undergo a phase transition at any temperature~\cite{wu,wang}. However, no phase transition can be expected even if $p_{\mathrm{A}}>0$, since then the system will behave like a ferrimagnet. Namely, starting from the ground state spin arrangement ($m_{\mathrm{A}},m_{\mathrm{B}},m_{\mathrm{C}})=(-p_{\mathrm{A}},1,1$), the increasing thermal excitations will make spins on the diluted sublattice {\it gradually} flip in the field direction with no sharp phase transition. Apparently, for $h/|J| > 6$ all spins are aligned to the field direction already at zero temperature, i.e. the system is in the paramagnetic state, and therefore no phase transition can occur as the temperature is increased. For this reason, in the following we will focus on the critical behavior of the system in the region $h/|J| \in \left[0,3\right]$.

\begin{table}[t!]
\caption{Ground-state sublattice magnetizations ($m_{\mathrm{A}},m_{\mathrm{B}},m_{\mathrm{C}}$) with the corresponding energies per site $\langle H \rangle/|J|N$, for different field $h/|J|$ intervals.}
\label{tab:GS}
\centering
\begin{tabular}{lccc}
\hline
$\frac{h}{|J|}$  & $\left(0,3\right]$  & $\left(3,6\right]$ & $(6,\infty)$ \\
\hline
($m_{\mathrm{A}},m_{\mathrm{B}},m_{\mathrm{C}}$) & ($p_{\mathrm{A}},1,-1$)  & ($-p_{\mathrm{A}},1,1$)  &  ($p_{\mathrm{A}},1,1$)\\
\hline
$\frac{\langle H \rangle}{|J|N}$ & $-1-\frac{hp_{\mathrm{A}}}{3}$   & $1-2p_{\mathrm{A}}-\frac{h(2-p_{\mathrm{A}})}{3}$   & $1+2p_{\mathrm{A}}-\frac{h(2+p_{\mathrm{A}})}{3}$ \\
\hline
\end{tabular}
\end{table}

\subsection{Critical and tricritical behavior}
\begin{figure}[t!]
\centering
    \includegraphics[scale=0.5]{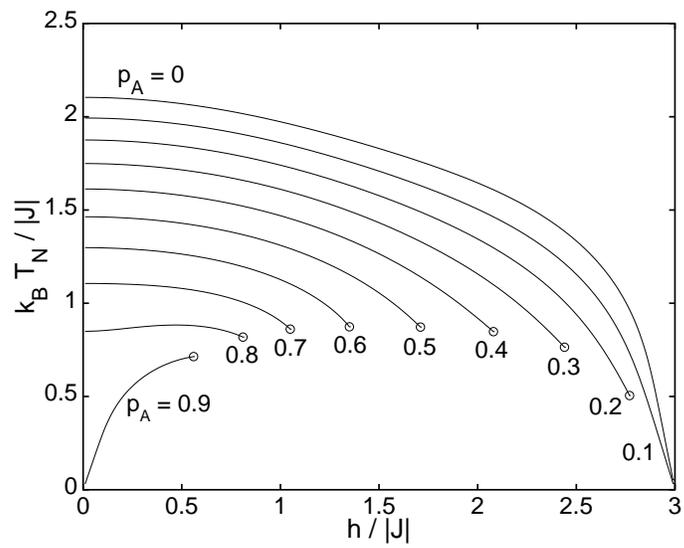}
\caption{Phase boundaries as functions of the field $h/|J|$ for different values of the concentration $p_{\mathrm{A}}$. The circles represent the tricritical points.}
\label{fig:PD_h-T}
\end{figure}

In order to investigate the selective dilution effects on the critical behavior, it is useful to start with the fully diluted case of $p_{\mathrm{A}}=0$, corresponding to the non-frustrated honeycomb lattice. Then, the increasing $p_{\mathrm{A}}$ can be looked upon as the decoration of the honeycomb lattice by randomly placing the magnetic ions into the centers of the hexagons (Fig.~\ref{fig:lattice}) and thus gradually increasing the frustration in the lattice. The effects on the critical behavior are apparent from Fig.~\ref{fig:PD_h-T}, in which the phase boundaries are plotted in the field-temperature plane for different values of $p_{\mathrm{A}}$. For $p_{\mathrm{A}}=0$, the system shows the phase transition within the field range $0 < h/|J| \le 3$, in accordance with the previous studies~\cite{wu,wang}, and the transition (or N\'{e}el) temperature at zero field $k_BT_N/|J|=2.1038$ coincides with the earlier EFT result for the honeycomb lattice~\cite{tagg}. With increasing $p_{\mathrm{A}}$ the transition temperatures tend to decrease, however, no qualitative changes occur as long as $p_{\mathrm{A}}$ is relatively small, e.g., $p_{\mathrm{A}}=0.1$. Nevertheless, upon further increase of $p_{\mathrm{A}}$, e.g., $p_{\mathrm{A}}=0.2$ and larger, at sufficiently high fields the order of the transition changes from the second to first one at a tricritical point (TCP). Furthermore, for $p_{\mathrm{A}} \gtrsim 0.8$ the character of the dependence of the critical temperature on the field changes from decreasing to increasing, and for $p_{\mathrm{A}} \approx 0.8$ even both tendencies can be observed in the same curve within different ranges of the field.

%  increases  on transition temperature more clearly, in figure \ref{fig:PD_h-T} the phase boundaries are plotted in the field-temperature plane, for different values of $p_{\mathrm{A}}$. We can observe that for small dilution the increasing field raises the critical temperature, while just the opposite is true at a higher degree of dilution. Hence, the effect of the increasing selective dilution here resembles the effect of the increasing field, seen in figure~\ref{fig:PD_p-T}, in the sense that as the dilution increases the critical behavior changes from that typical for a frustrated system to that typical for a non-frustrated case. Nevertheless, besides relieving the frustration, the increasing dilution effectively decreases the system's total coordination number $z$ from the non-diluted (triangular lattice) value of 6 to the totally diluted (honeycomb lattice) value of 3, thus pushing the critical temperature to lower values and ultimately to zero at $h/|J|=z$. The combined effect of the above factors can result in a rather intricate behavior of the phase boundaries. For example, for some intermediate values of $p_{\mathrm{A}} \approx 0.8$, as the field increases the initial increase in $T_N$ is followed by some decrease and this pattern is repeated once more.

\begin{figure}
\centering
    \includegraphics[scale=0.5]{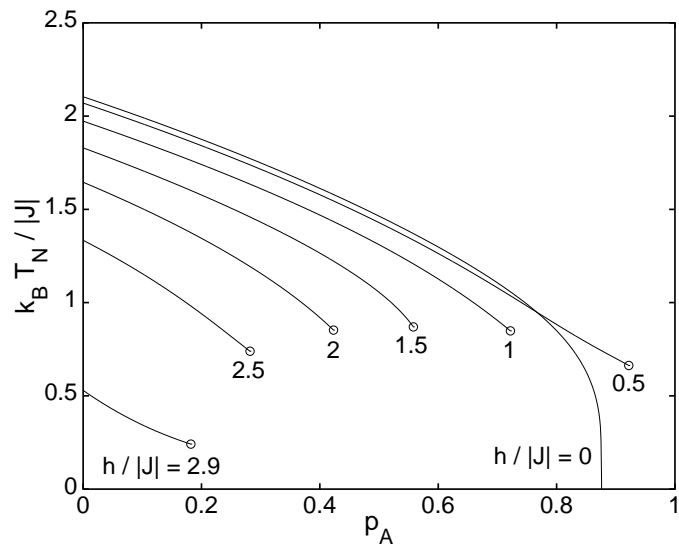}
\caption{Phase boundaries as functions of the concentration $p_{\mathrm{A}}$ for different values of the field $h/|J|$. The circles represent the tricritical points.}
\label{fig:PD_p-T}
\end{figure}
\begin{figure}
\centering
    \includegraphics[scale=0.5]{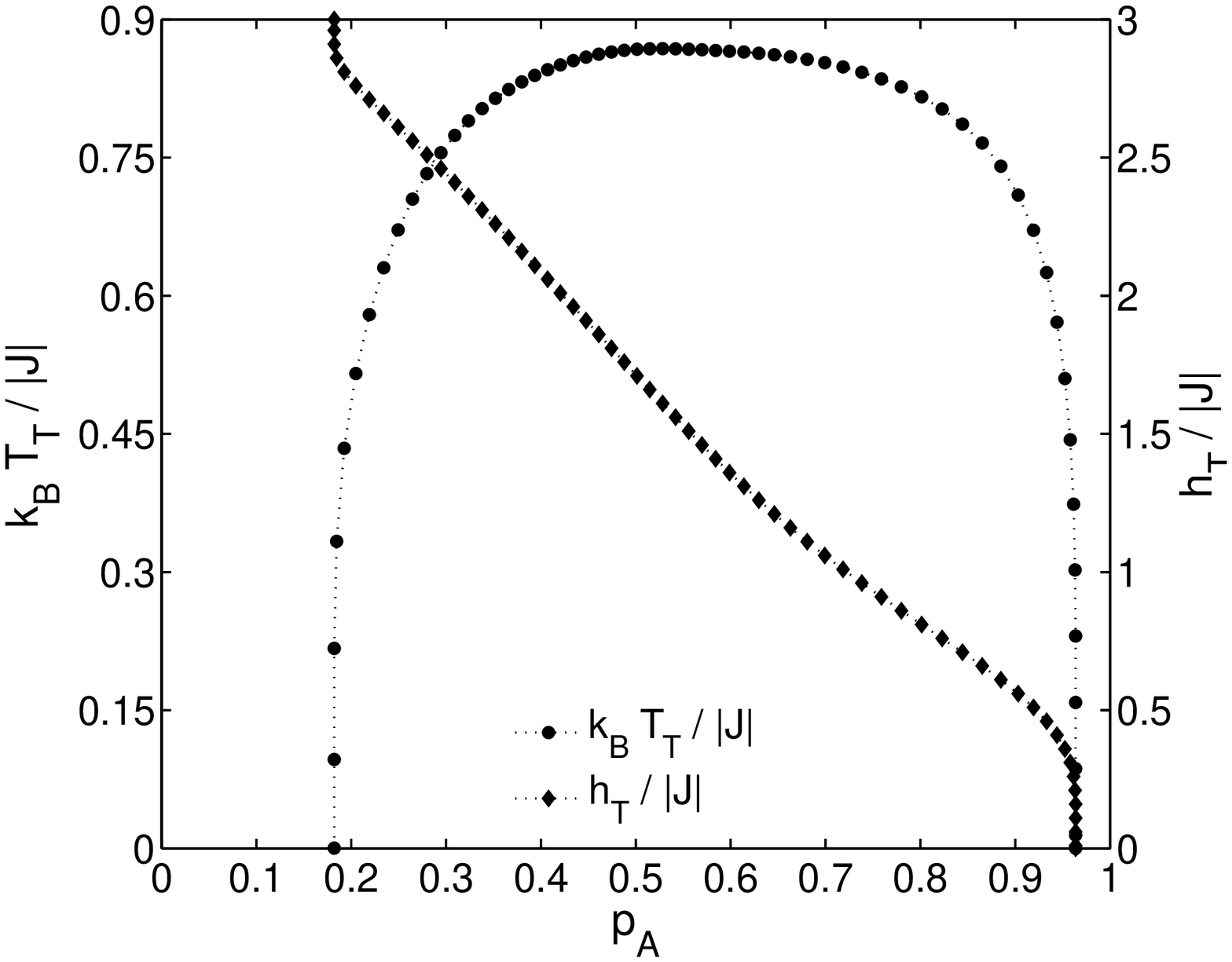}
\caption{Tricritical temperature and tricritical field as functions of the concentration $p_{\mathrm{A}}$.}
\label{fig:TCP}
\end{figure}

Phase diagrams in the concentration-temperature parameter space are presented in Fig.~\ref{fig:PD_p-T}, for several values of the field $h/|J|$. For $h/|J|=0$, a long-range order emerges below $p_{\mathrm{A}}^c=0.8753$ and the critical temperature increases with the decreasing value of $p_{\mathrm{A}}$ up to $k_BT_N/|J|=2.1038$ for $p_{\mathrm{A}}=0$, corresponding to the honeycomb lattice. In a finite field, the phase transition appears to take place for all concentrations, however, above a certain value of $p_{\mathrm{A}}$ the order of the transition changes to the first order at a TCP. With the increasing field the second-order transition temperatures decrease, as well as the concentrations $p_{\mathrm{A}}$ corresponding to the TCPs. 

Hence, from the point of view of the varying degree of frustration, the following statements can be formulated. Relatively little frustrated systems with small values of $p_{\mathrm{A}} < 0.1821$ show only the second-order transitions. For more frustrated systems, within $0.1821 \leq p_{\mathrm{A}} \leq 0.9638$, the transitions remain second order at low fields but they change at TCPs to first order for higher fields. Finally, for highly frustrated cases, with $p_{\mathrm{A}} > 0.9638$, there are no TCPs and the transitions are of first order at any field. In order to better visualize the behavior of both the tricritical temperature and tricritical field with the varying $p_{\mathrm{A}}$, we plot these dependencies in a separate Fig.~\ref{fig:TCP}. The tricritical behavior within $0.1821 \leq p_{\mathrm{A}} \leq 0.9638$ is characterized by the monotonically decreasing tricritical field from $h_T/|J|=3$ at $p_{\mathrm{A}} = 0.1821$ down to $h_T/|J|=0$ at $p_{\mathrm{A}} = 0.9638$. On the other hand, the tricritical temperature shows an inverted-U-shape dependence with the highest values of $k_BT_T/|J| = 0.8682$ in the middle and approaching zero at both ends of the interval $0.1821 \leq p_{\mathrm{A}} \leq 0.9638$.

\subsection{First-order phase transitions}
\begin{figure}[t!]
\centering
    \subfigure[$h/|J|=0.5,p_{\mathrm{A}}=1$]{\includegraphics[scale=0.33]{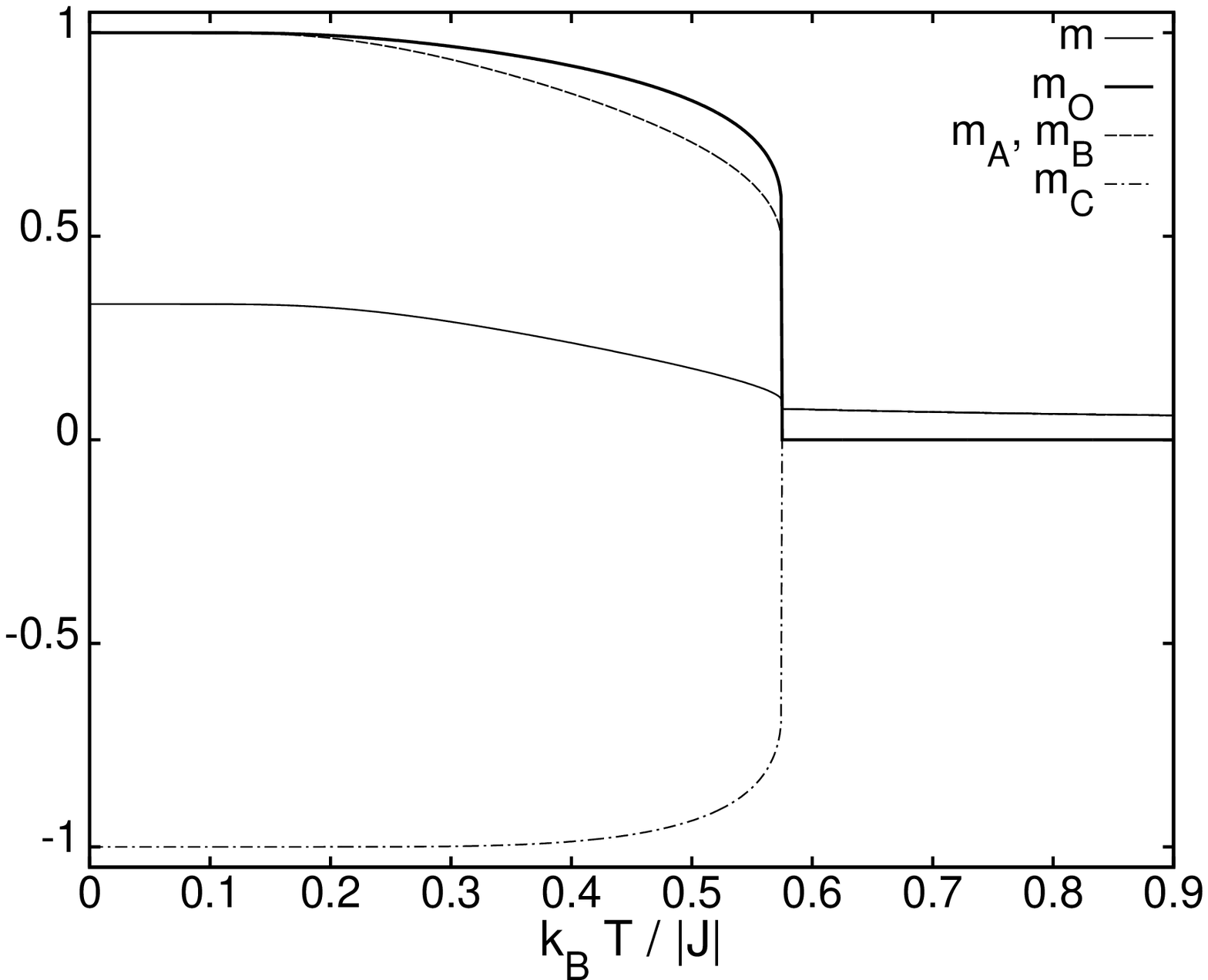}\label{fig:T_m_h05_p1}}
    \subfigure[$h/|J|=0.5,p_{\mathrm{A}}=0.5$]{\includegraphics[scale=0.33]{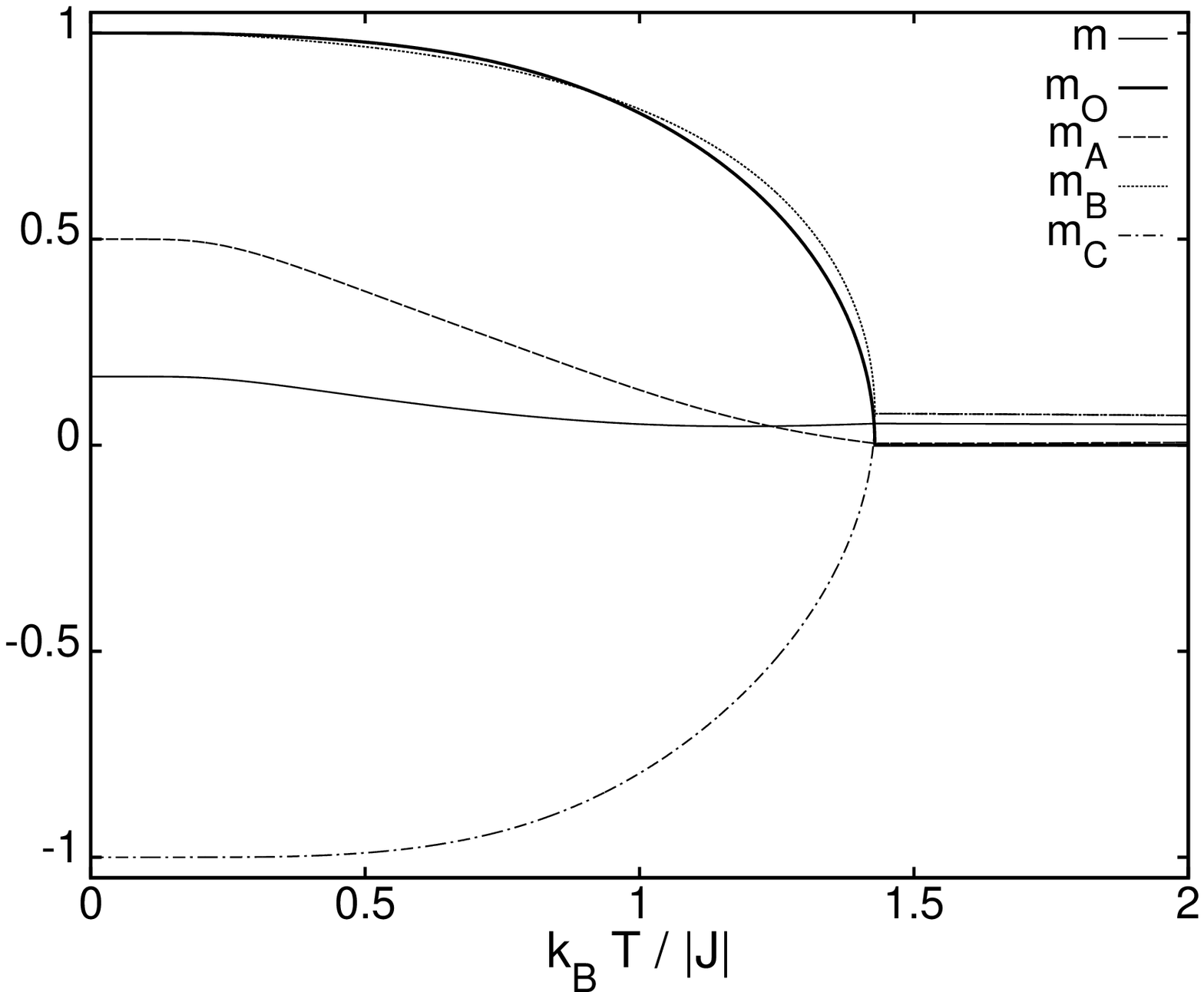}\label{fig:T_m_h05_p05}}\\
    \subfigure[$h/|J|=2.5,p_{\mathrm{A}}=1$]{\includegraphics[scale=0.33]{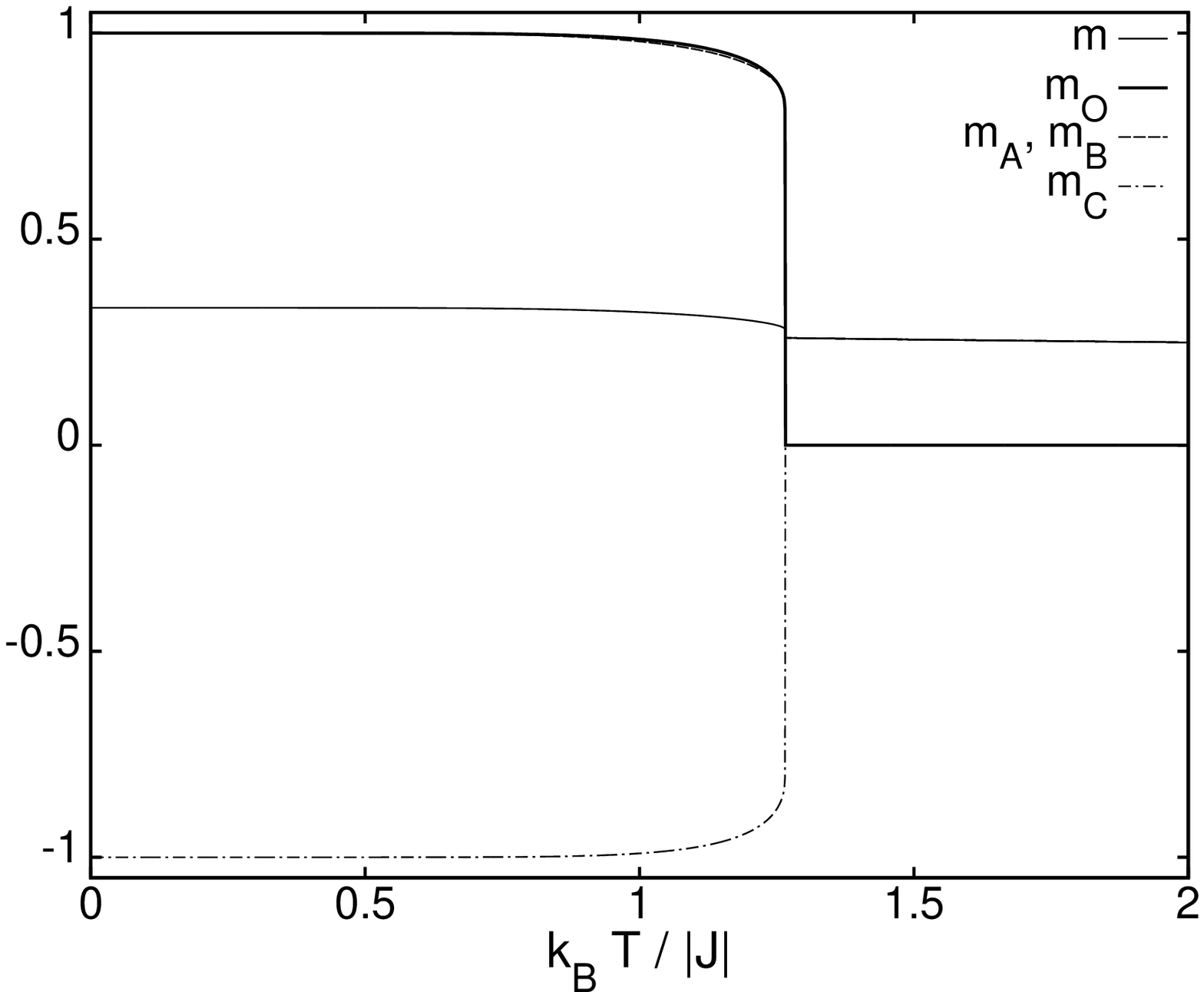}\label{fig:T_m_h25_p1}}
    \subfigure[$h/|J|=2.5,p_{\mathrm{A}}=0.5$]{\includegraphics[scale=0.33]{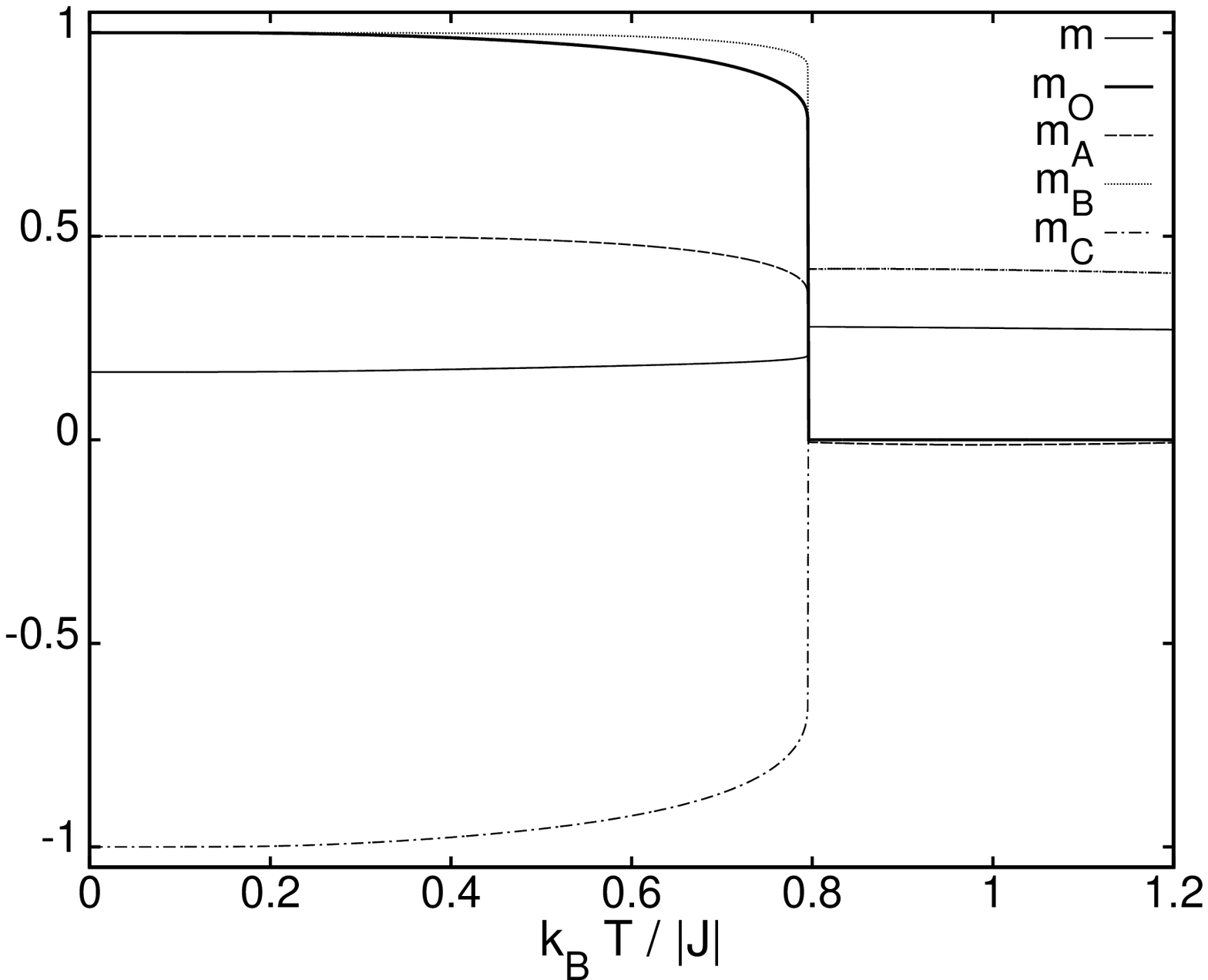}\label{fig:T_m_h25_p05}}
%    \subfigure[$h/|J|=4,p_{\mathrm{A}}=0.7$]{\includegraphics[scale=0.35]{T_mag_H4_P07.eps}\label{fig:T_mag_H4_P07}}
\caption{Temperature variations of the sublattice magnetizations $m_{\mathrm{A}}$, $m_{\mathrm{B}}$ and $m_{\mathrm{C}}$, the total magnetization $m$, and the order parameter $m_{\mathrm{O}}$, for different values of $h/|J|$ and $p_{\mathrm{A}}$.}
\label{fig:finite_field_m}
\end{figure}

\begin{figure}[t!]
\centering
    \includegraphics[scale=0.5]{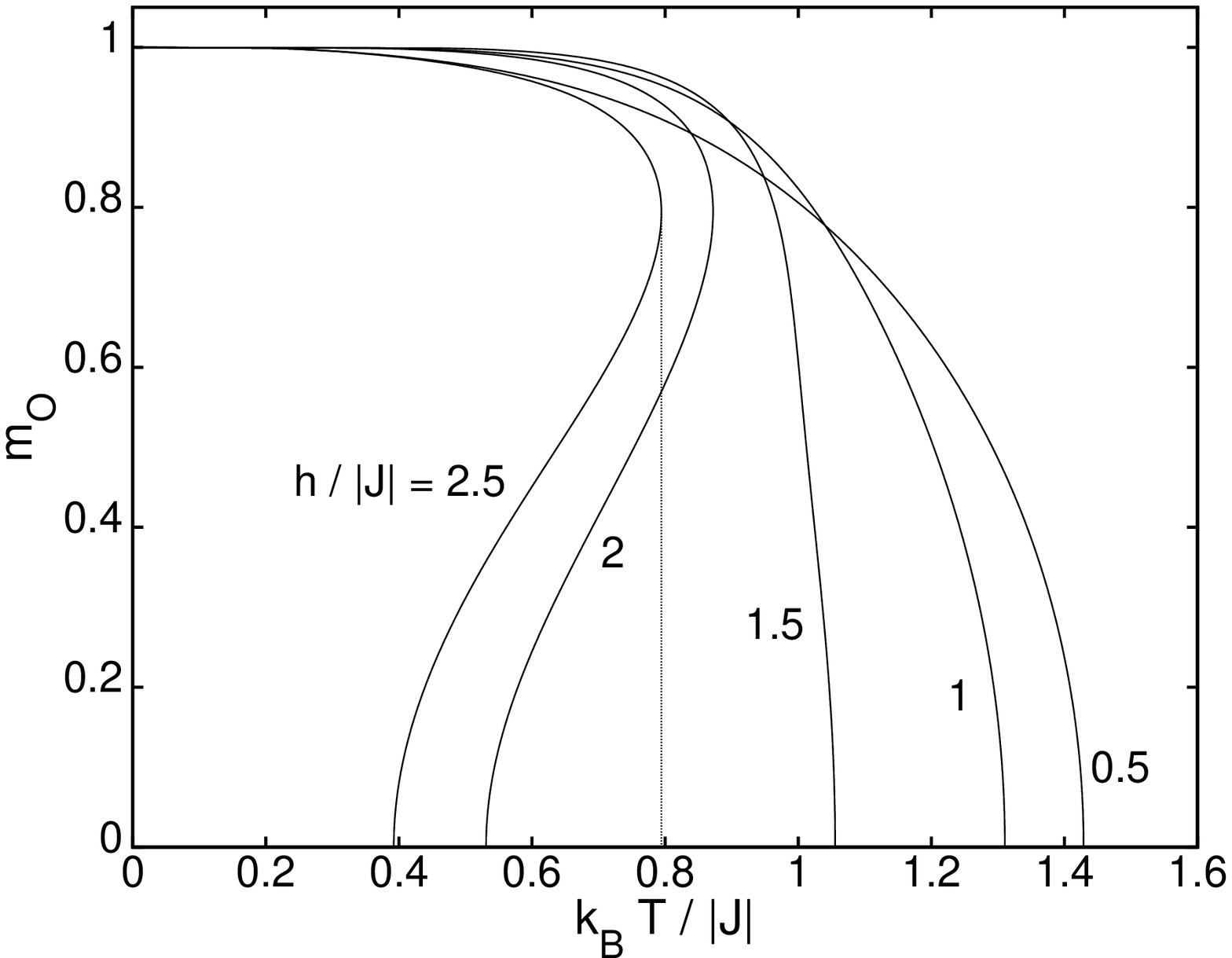}
\caption{Temperature variation of the order parameter $m_{\mathrm{O}}$ for $p_{\mathrm{A}}=0.5$ and different values of the field $h/|J|$.}
\label{fig:T_o_P05}
\end{figure}

Above, we established a line of tricritical temperatures below which the transitions are of first order. Even though the EFT formalism is able to determine the first-order character of the transition, unfortunately, it lacks an expression for the free energy and thus tools for its localizing, i.e., determining the first-order transition boundary. Nevertheless, typical first-order transition features can be observed in the behavior of various quantities. In Fig.~\ref{fig:finite_field_m} we plot temperature dependencies of the sublattice magnetizations, $m_{\mathrm{A}},m_{\mathrm{B}},m_{\mathrm{C}}$, the total magnetization, $m=(m_{\mathrm{A}}+m_{\mathrm{B}}+m_{\mathrm{C}})/3$, and the order parameter, $m_{\mathrm{O}}$, in different regions of the parameter space. Apparently, there are qualitative difference among different cases. Indeed, while the curves in Fig.~\ref{fig:T_m_h05_p05}, corresponding to a relatively low field and considerable dilution, vanish to zero continuously, rather sharp drops typical for the first-order transitions are observed in the curves corresponding to either high fields (Figs.~\ref{fig:T_m_h25_p1} and \ref{fig:T_m_h25_p05}) or a relatively low field but no dilution (Fig.~\ref{fig:T_m_h05_p1}). More detailed computations in a narrow region around the point where the curves appear to be discontinuous reveal that the vertical drop results from the fact that above a certain value of the field the corresponding quantities cease to be single-valued functions of the temperature. By exploring other solutions, we find the formation of wiggles, like those shown in Fig.~\ref{fig:T_o_P05} for the order parameter, with $h/|J|=2$ and $2.5$. Within effective-field approaches such a behavior signals appearance of a first-order transition~\cite{kinc,zuko1}.
  
At this point we note that our EFT results suggest that the pure system ($p_{\mathrm{A}}=1$) shows first-order transitions for all fields. However, the pure system in a field was argued to belong to the same universality class as the three-state Potts model, which displays a second-order phase transition in two dimensions~\cite{schick}, and the transition appeared second order also within the Monte Carlo mean-field approach~\cite{netz}. Therefore, it is legitimate to ask whether the first-order transitions are not just an artifact of the EFT approximation. Indeed, there are examples of other spin systems, such as some mixed-spin systems~\cite{zuko2,buen}, for which EFT predicted the tricritical behavior but it was not confirmed by more accurate methods.

\begin{figure}[t!]
\centering
    \includegraphics[scale=0.5]{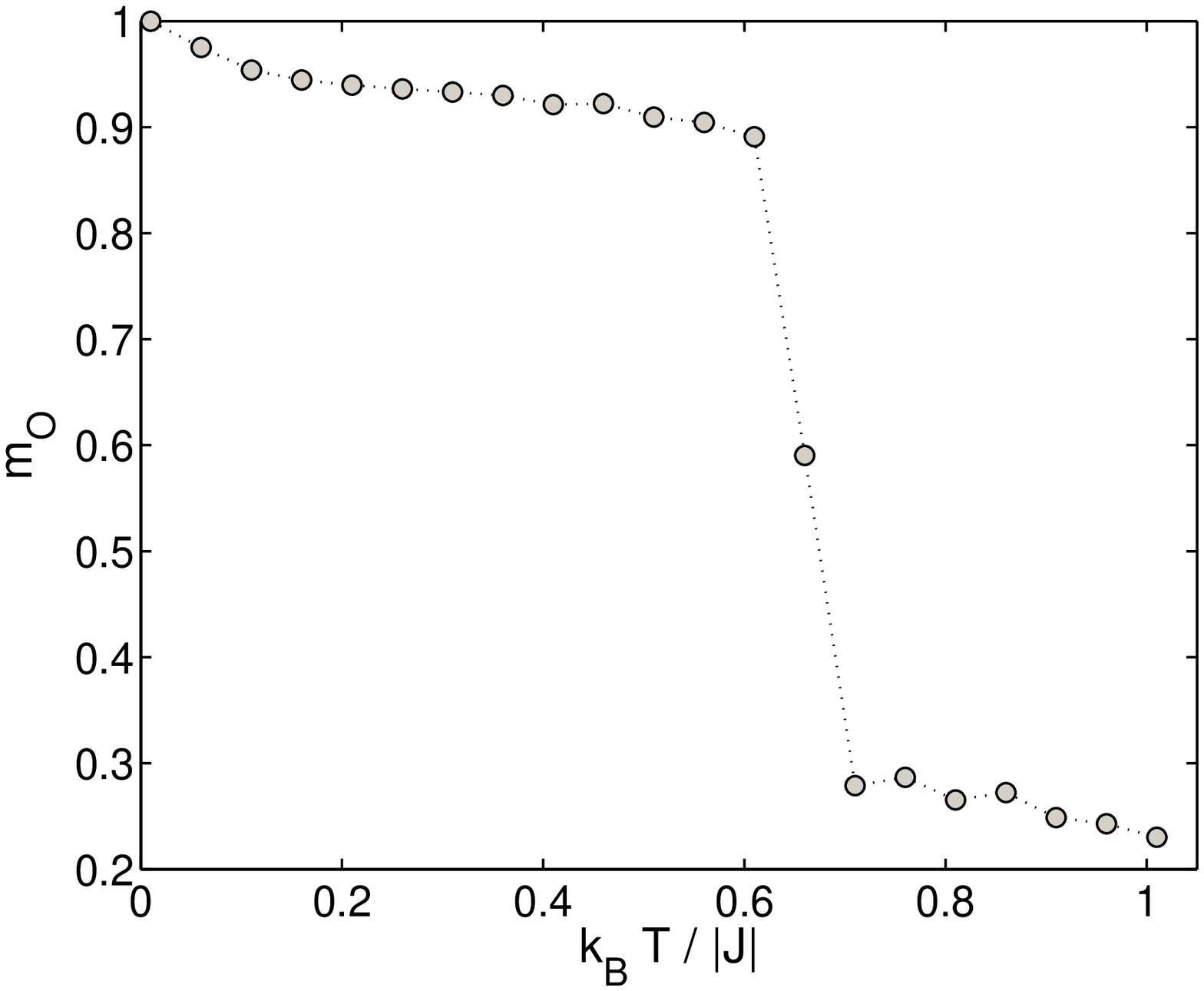}
\caption{Temperature variation of the order parameter $m_{\mathrm{O}}$ for the increasing temperature, for $p_{\mathrm{A}}=0.5$, $h/|J|=2.95$ and MCS$=2 \times 10^4$.}
\label{fig:mo-T}
\end{figure}

\begin{figure}[t!]
\centering
    \subfigure[Distribution of $m_{\mathrm{O}}$ in (pseudo)critical temperature]{\includegraphics[scale=0.5]{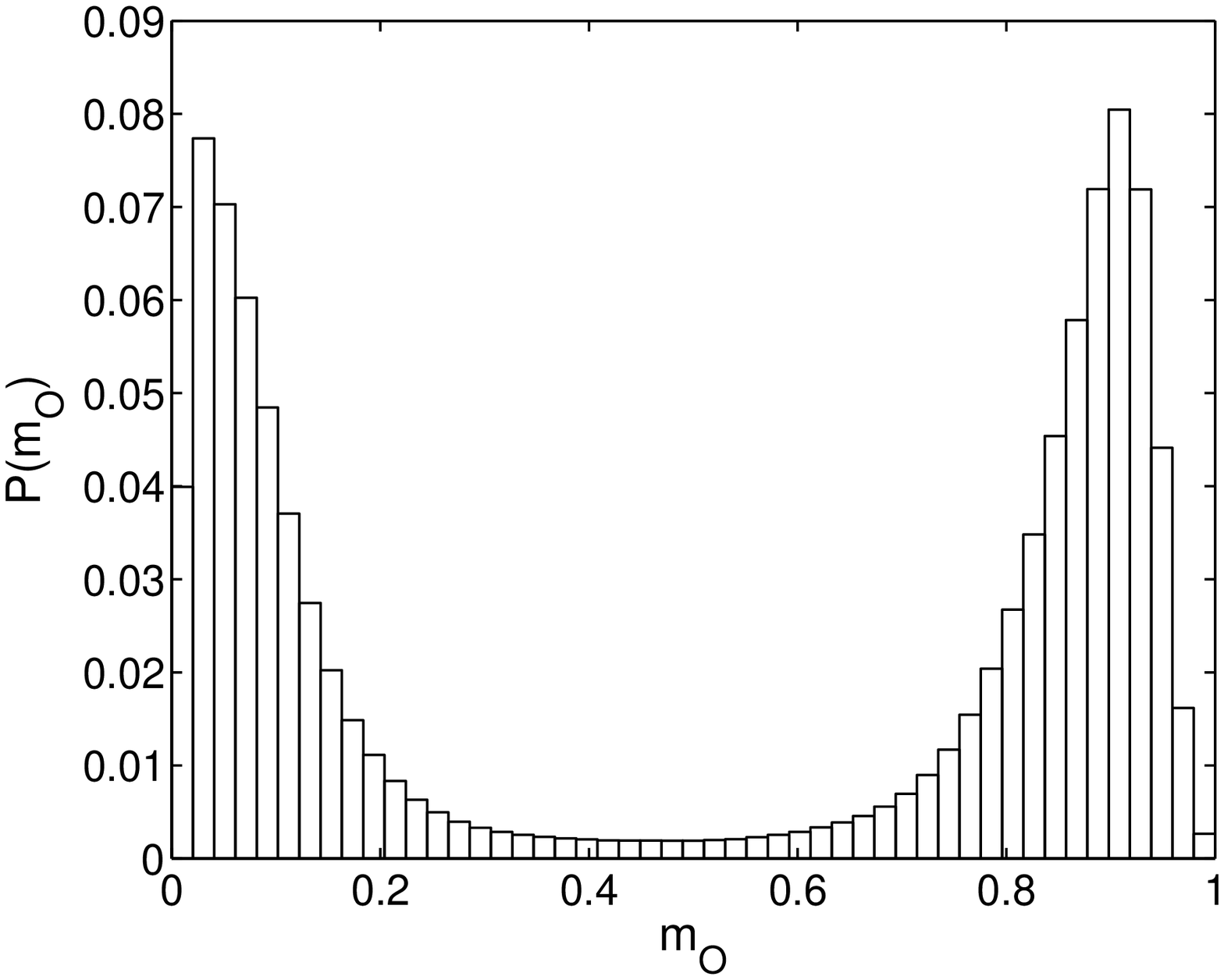}\label{fig:mo-hist}}
    \subfigure[Time evolution of $m_{\mathrm{O}}$ in (pseudo)critical temperature]{\includegraphics[scale=0.5]{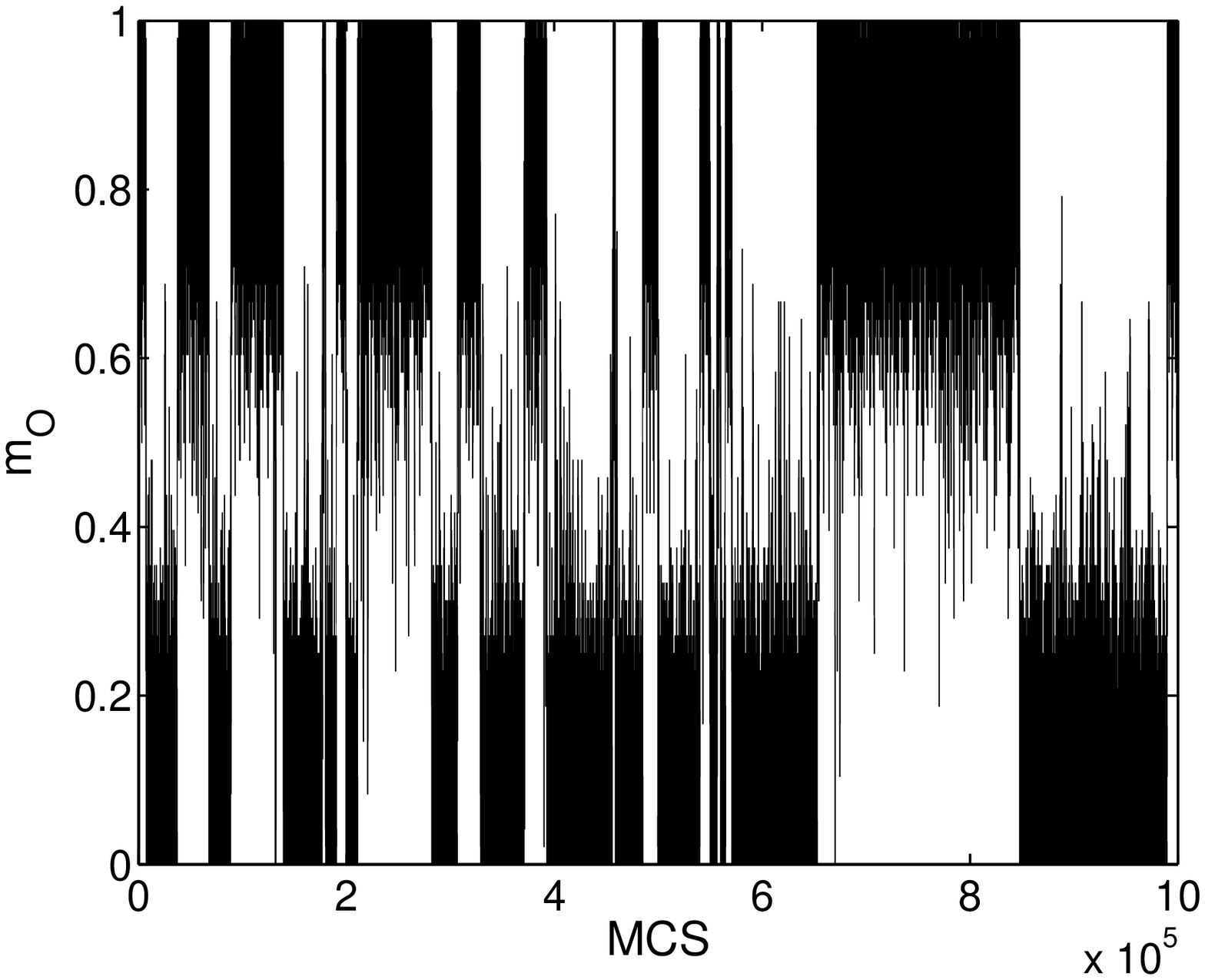}\label{fig:mo-mcs}}
\caption{MC simulation results for the order parameter $m_{\mathrm{O}}$ in the (pseudo)critical temperature $k_BT_{N}/|J|=0.6545$, for $L=12$, $p_{\mathrm{A}}=0.5$ and $h/|J|=2.95$. Figure (a) shows the distribution of the entire history of $9 \times 10^6$ MCS and (b) shows the time evolution of the first $10^6$ MCS, after equilibration over the initial $10^6$ MCS.}
\end{figure}

In order to confirm the existence of the first-order transition in the present system, we additionally performed Monte Carlo simulations~\cite{land00}. We note, however, that generally this is not an easy task since the first-order transition features often show up only when very large system sizes are used, which for highly frustrated systems means enormous equilibration times. For example, in our tests for $p_{\mathrm{A}}=1$ and $h/|J|=2.95$ even as many as ten million Monte Carlo steps per spin (MCS) was not enough to reach equilibrium for the linear lattice sizes $L \sim 200$. Therefore, to demonstrate the existence of the first-order transition, we choose a case in which frustration is partially relieved, namely the system with $p_{\mathrm{A}}=0.5$, and for which the EFT predicts a first-order transition (see Figs.~\ref{fig:PD_h-T} and~\ref{fig:TCP}). For this case, the relaxation times are reasonable and a fist-order transition character is observed at already as small linear lattice sizes as $L=12$. In our MC simulations the number of MCS ranged from $2 \times 10^4$, in the case of obtaining temperature dependences of various quantities, up to $10^7$, in the case of evaluating distributions at a fixed temperature. First 10\% of MCS were discarded for thermalization and the remaining were taken for thermal averaging. In Fig.~\ref{fig:mo-T} we show a temperature dependence of the order parameter for $p_{\mathrm{A}}=0.5$ and $h/|J|=2.95$, for the increasing temperature. The curve displays a typical discontinuous first-order transition behavior with the jump above $k_BT/|J| \approx 0.6$. A more precise (pseudo)critical temperature $k_BT_N(L=12)/|J|=0.6545$ is found by tunning the temperature so that the order parameter displays a bimodal distribution with equally high peaks (Fig.~\ref{fig:mo-hist}). The time evolution of the order parameter, shown for the first $10^6$ MCS in Fig.~\ref{fig:mo-mcs}, demonstrates that at $k_BT_N(L=12)/|J|$ the system switches between the ordered ($m_{\mathrm{O}} > 0$) and disordered ($m_{\mathrm{O}} = 0$)\footnote{Due to the finiteness of the lattice size in MC simulations the order parameter remains non-zero even in the disordered phase and the zero value is only achieved for $L \rightarrow \infty$.} states. The true critical temperature can be usually estimated from a finite-size scaling (FSS) analysis~\cite{ferr88}. Unfortunatelly, for frustrated spins systems increasing the lattice size would mean an enormous increase of the tunneling times needed to switch between the ordered and disordered states and thus hampering a reliable determination of the (pseudo)critical points needed for FSS by the equal height condition. We note that, although in our case we were able to observe clearly the bimodal distribution of the order parameter already at $L=12$, relatively small lattice sizes generally suppress the typical first-order transition features. For example, when we used much larger lattice sizes, such as $L=240$, some faint signs of the bimodal distribution appeared already for $h/|J|=2.5$. However, as mentioned above, for such lattice sizes enormous equilibration and tunneling times would be required to reliably confirm the transition order. Thus, localizing a TCP, in which the order of the transition changes from the second to the first one, by the Monte Carlo technique would require using sufficiently large lattice sizes along with some more sophisticated method, such as the multicanonical Monte Carlo method, to keep the tunneling times in reasonable limits, and then performing careful finite-size scaling analysis~\cite{ferr88}. Consequently, establishing the entire phase diagram, including the first-order transition boundaries and the tricritical points, in a broad field-concentration-temperature parameter space by Monte Carlo simulations is a computationally very demanding task and it is out of scope of the present study.

\section{Summary and conclusions}
We studied the critical behavior of a geometrically frustrated selectively site-diluted triangular lattice Ising antiferromagnet in a field within the framework of an effective-field theory with correlations. The selective dilution was carried out by random removal of magnetic ions from one of the three sublattices. In particular, we focused on the effects of the frustration-relieving selective dilution on the resulting phase diagram. In accordance with some previous studies, in zero field such a controlled dilution relieved geometrical frustration and resulted in a long-range ordering in the remaining two sublattices below some threshold concentration of magnetic sites. In finite fields the system was found to display the phase transition at any concentration, however, the order of the transition depended on its value. In particular, for relatively little frustrated systems with small values of the concentration of magnetic sites we found only the second-order transition at any field. For more frustrated systems, corresponding to larger values of the concentration, the system displayed a tricritical behavior in which the transition remained second-order at low fields but it changed to first order at higher fields. Finally, for highly frustrated cases, with the concentrations approaching the pure system, the transitions were identified as first order at any field. The existence of the first-order transitions in the region of moderate frustration and high fields was confirmed by Monte Carlo simulations.

Nevertheless, the EFT approach only produced the second-order transition boundaries and may have overestimated the region in the parameter space in which the first-order transitions are present. Therefore, additionally, it would be desirable to employ some other more reliable techniques to establish the first-order transition lines and also to check the extent of the parameter space in which they take place.

%To establish the transition order and the phase boundaries in a broad parameter space more reliably, . Such a study could also include the limiting case of the non-frustrated antiferromagnet on the honeycomb lattice (i.e., $p_{\mathrm{A}}=0$ case), the phase boundary of which was already determined~\cite{wu,wang} but the transition order does not seem to have been looked into.    

\section*{Acknowledgments}
This work was supported by the Scientific Grant Agency of Ministry of Education of Slovak Republic (Grant No. 1/0234/12). The authors acknowledge the financial support by the ERDF EU (European Union European regional development fund) grant provided under the contract No. ITMS26220120005 (activity 3.2.).
\let\thefigureSAVED\thefigure
\let\thetableSAVED\thetable

\newpage

\appendix
\section{}
List of coefficients $K_i$ ($i = 0-6$) and $L_j$ ($j = 0-15$) in Eqs.~\eqref{sub_mag_honey2}:
\begin{equation}
\begin{array}{l}
		\label{coeffsK}
				K_0 = p_{\mathrm{A}} a^6 \tanh \left( x + \beta h \right) |_{x=0}, \\
				K_1 = 3 p_{\mathrm{A}} a^5 b \tanh \left( x + \beta h \right) |_{x=0}, \\
				K_2 = 3 p_{\mathrm{A}} a^4 b^2 \tanh \left( x + \beta h \right) |_{x=0}, \\
				K_3 = p_{\mathrm{A}} a^3 b^3 \tanh \left( x + \beta h \right) |_{x=0}, \\
				K_4 = 3 p_{\mathrm{A}} a^2 b^4 \tanh \left( x + \beta h \right) |_{x=0}, \\
				K_5 = 3 p_{\mathrm{A}} a b^5 \tanh \left( x + \beta h \right) |_{x=0}, \\
				K_6 = p_{\mathrm{A}} b^6 \tanh \left( x + \beta h \right) |_{x=0},			
\end{array}
\end{equation}
and
\begin{equation}
\begin{array}{l}
		\label{coeffsL}
				L_0 = a_{\mathrm{A}}^3 a^3 \tanh \left( x + \beta h \right) |_{x=0}, \\
				L_1 = 3 a_{\mathrm{A}}^2 a^3 b  \tanh \left( x + \beta h \right) |_{x=0}, \\
				L_2 = 3 a_{\mathrm{A}}^3 a^2 b  \tanh \left( x + \beta h \right) |_{x=0}, \\
				L_3 = 3 a_{\mathrm{A}} a^3 b^2  \tanh \left( x + \beta h \right) |_{x=0}, \\
				L_4 = 3 a_{\mathrm{A}}^3 a b^2  \tanh \left( x + \beta h \right) |_{x=0}, \\
				L_5 = 9 a_{\mathrm{A}}^2 a^2 b^2  \tanh \left( x + \beta h \right) |_{x=0}, \\
				L_6 = a^3 b^3  \tanh \left( x + \beta h \right) |_{x=0}, \\
				L_7 = a_{\mathrm{A}}^3 b^3  \tanh \left( x + \beta h \right) |_{x=0}, \\
				L_8 = 9 a_{\mathrm{A}} a^2 b^3  \tanh \left( x + \beta h \right) |_{x=0}, \\
				L_9 = 9 a_{\mathrm{A}}^2 a b^3  \tanh \left( x + \beta h \right) |_{x=0}, \\
				L_{10} = 9 a_{\mathrm{A}} a b^4  \tanh \left( x + \beta h \right) |_{x=0}, \\
				L_{11} = 3 a^2 b^4  \tanh \left( x + \beta h \right) |_{x=0}, \\
				L_{12} = 3 a_{\mathrm{A}}^2 b^4  \tanh \left( x + \beta h \right) |_{x=0}, \\
				L_{13} = 3 a b^5 \tanh \left( x + \beta h \right) |_{x=0}, \\
				L_{14} = 3 a_{\mathrm{A}} b^3  \tanh \left( x + \beta h \right) |_{x=0},	\\
				L_{15} =  b^6 \tanh \left( x + \beta h \right) |_{x=0}.			
\end{array}
\end{equation}

The coefficients can be calculated by applying a mathematical relation $\exp({\lambda D})f(x)=f(x+\lambda)$.

\let\thefigure\thefigureSAVED
\let\thetable\thetableSAVED

%\section*{References}
\bibliographystyle{elsarticle-num}
\biboptions{sort&compress}

\end{document}